# DFT study of optical properties of pure and doped Graphene

**Pooja Rani[1], Girija S Dubey[2] and V. K. Jindal[1*]**

[1]Department of Physics, Panjab University Chandigarh-160014, India.

[2] Department of Earth & Physical Sciences, York College of City University of New York, Jamaica, NY11451

*Ab-initio* calculations based on density functional theory (DFT) have been performed to study the optical properties of pure graphene and have been compared to that of individual boron (B), nitrogen (N) and BN co-doped graphene sheet. The effect of doping has been investigated by varying the concentrations of dopants from 3.125 % (one atom of the dopant in 32 host atoms) to 6.25 % (six dopant atoms in 50 host atoms) for individual B and N doping and from 37.5 % (one B/N pair in 32 host atoms) to 18.75 % for BN co-doping. Positions of the dopants have also been varied for the same concentration of substitution doping. The dielectric matrix has been calculated within the random phase approximation (RPA) using VASP (Vienna *ab-initio* Simulation Package) code. The dielectric function, absorption spectrum and energy loss-function of single layer graphene sheet have been calculated for light polarization parallel and perpendicular to the plane of graphene sheet and compared with doping graphene. The calculated dielectric functions and energy-loss spectra are in reasonable agreement with the available theoretical and experimental results for pure graphene. It has been found that individual B and N doping does not significantly affect the imaginary dielectric function and hence the absorption spectra. However, significant red shift in absorption towards visible range of the radiation at high doping is found to occur for the B/N co-doping. The results can be used to tailor the optical properties of graphene in visible region.

**Keywords**- Graphene, density functional theory, optical properties, absorption spectra

## 1. Introduction

The specific electronic structure of a single layer of graphite, which is called graphene [1], gave rise to intense investigations of optoelectronic properties of graphene-based materials [2-6]. For example, it can be optically contrasted from the substrate, despite being only a single atom thick [7, 8]. It can also lead to luminescence through chemical and physical treatments [9-11]. These properties make it an ideal photonic and optoelectronic material [2]. The rise in interest of graphene in photonics and optoelectronics is shown by its applications ranging from solar cells and light-emitting devices to touch screens, photo-detectors and ultrafast lasers. This is because the combination of its unique optical and electronic properties can be fully exploited in nano-photonics.

Due to its visual transparency graphene has a potential as transparent coatings. Optical absorption of graphene is anisotropic for light polarization being parallel or perpendicular to the axis normal to the sheet. Experiments have shown that as compared to graphite the optical and loss spectra of graphene exhibits a redshift of absorption bands and $\pi + \sigma$ electron plasmon and disappearance of bulk plasmons [4, 6]. Optical properties are the prominent characteristics that distinct it from graphite. It was shown by T. Ebernil et. al [5] that $\pi$ and $\pi+\sigma$ surface plasmon modes in free-standing single sheets at 4.7 and 14.6 eV, which are substantially red-shifted from their values in graphite.

Among the many areas in which graphene may excel is, e.g., its application for sensors due to the sensitivity of its electronic structure to adsorbates. Low loss energy electron spectroscopy provides a way of detecting changes in the electronic structure, which are highly spatially resolved.

Optical properties of graphite and graphene have been extensively studied by and Sedelnikova et al. [12] and Marinopoulos et al. [2, 13]. Marinopoulos et al have calculated the absorption spectrum for different values of c/a ratio and compared it with that of BN sheet. Eberlein at al. [5] carried out the experimental spectroscopy of graphene in conjunction with *ab- initio* calculations of the low loss function. Sedelnikova et al. [12] have studied the effects of ripples on optical properties of graphene. But no systematic study using the *ab-initio* methods to find the effect of hetero-atom doping on the optical properties of graphene has been reported to the best of our knowledge so far.

Since graphene in pure form is transparent in the visible part of the spectrum, so in order for graphene based optoelectronic devices to be useful, it is beneficial if they can be tailored to absorb specific wavelength region of the spectra. In the present work we plan to carry out investigations on the effect of B, N and BN co-doping at different concentrations on the absorption spectrum and other optical parameters of graphene with the speculation in mind of role of doping in tailoring the absorption wavelength region.

## 2. Theory and Computational details

For the present analysis, VASP (Vienna *Ab-initio* Simulation Package) [14, 15] code based on density functional theory (DFT) was used. The approach is based on an iterative solution of the Kohn-Sham equation [16] of the density functional theory in a plane-wave set with the projector-augmented wave pseudopotentials. We adopted the Perdew-Burke-Ernzerhof (PBE) [17] exchange-correlation (XC) functional of the generalized gradient approximation (GGA) in our calculations. The plane-wave cutoff energy was set to 400 eV. The $4 \times 4$ supercell (consisting of 32 atoms) has been used to simulate the isolated sheet and the sheets are separated by larger than 12 Å along the perpendicular direction to avoid interlayer interactions. The Monkhorst-Pack scheme is used for sampling the Brillouin zone. In the calculations, the structures are fully relaxed with a Gamma–centred $7 \times 7 \times 1$ k-mesh. The partial occupancies were treated using the tetrahedron methodology with Blöchl corrections [18]. For geometry optimizations, all the internal coordinates were relaxed until the Hellmann-Feynman forces were less than 0.005 Å. Dielectric function $\varepsilon\ (\omega)$ was calculated in the energy interval from 0 to 25 eV.

To calculate the optical properties, we used DFT within the Random Phase Approximation (RPA) approach in which the local field effects are included at Hartree level only. Only interband transitions are taken into account, so there may be inaccuracy in dielectric function at low energies. For a translationally invariant system, the Fourier transform of the frequency dependent symmetric dielectric matrix in the RPA is given by [19]

$$\varepsilon_{G,G'}(q, \omega) = \delta_{G,G'}(q, \omega) - \frac{4\pi e^2}{|G + q||G' + q|} \chi^0_{G,G'}(q, \omega)$$

where $G$ and $G'$ are reciprocal lattice vectors and q stands for the Bloch vector of the incident wave. The matrix $\chi^0(q, \omega)$ is the irreducible polarizability matrix in case of the independent particle derived by Adler and Wiser [20, 21] in the context of the self-consistent field approach.

The dielectric function can be written as sum of real and imaginary part i. e. $\varepsilon = \varepsilon' + i\ \varepsilon''$ and for calculating both these components for graphene in the present work, different polarizations of electric field w.r.t to the $c$ axis (which is normal to the plane of graphene sheet) are taken into account. That means $\varepsilon$ is calculated both for in-plane light polarization ($E \perp c$) and out of plane polarization or parallel to the $c$ axis ($E//c$).

VASP calculates the frequency dependent dielectric matrix after the electronic ground state has been determined. The imaginary part is determined by a summation over empty states using the equation:

$$\epsilon''_{\alpha\beta}(\omega) = \frac{4\pi^2 e^2}{\Omega} Lim_{q \to 0} \frac{1}{q^2} \sum_{c,v,k} 2\omega_k \delta(\epsilon_{ck} - \epsilon_{vk} - \omega)\langle u_{ck+e_\alpha q}|u_{vk}\rangle \langle u_{ck+e_\beta q}|u_{vk}\rangle$$

where the indices $\alpha$ and $\beta$ are he Cartesian components, vectors $e_\alpha$ and $e_\beta$ are the unit vectors along three directions, c and v refer to conduction and valence band states respectively, $\epsilon_{ck}$ refers to energy of conduction band and $\epsilon_{vk}$ refers to energy of valence band and $u_{ck}$ is the cell periodic part of the orbitals at the k-point $k$.

The real part of the dielectric tensor $\epsilon'$ is obtained by the usual Kramers- Kronig transformation

$$\epsilon'_{\alpha\beta}(\omega) = 1 + \frac{2}{\pi} P \int_0^\infty \frac{\epsilon''_{\alpha\beta}(\omega')\omega'}{\omega'^2 - \omega^2 + i\eta} d\omega'$$

where $P$ denotes the principle value. The method is explained in detail in Ref. [19]

The electron energy-loss spectrum (EEL) function was given by the imaginary part of $\left(\frac{1}{\varepsilon_{\alpha,\beta(\omega)}}\right)$, .given by $\frac{\varepsilon''_{\alpha\beta}}{\varepsilon''_{\alpha\beta}{}^2 + \varepsilon'_{\alpha\beta}{}^2}$.

Reflectivity is calculated by using the expression

$$R = \frac{(n-1)^2 + k^2}{(n+1)^2 + k^2}$$

where $n$ is real part of refractive index and $k$ is imaginary part of the refractive index.

## 3. Results and Discussion

## 3.1 Optical properties of pure graphene

The imaginary part of the dielectric function of pure graphene obtained in this way, which directly gives the absorption spectra of graphene, is presented in Fig. 1.

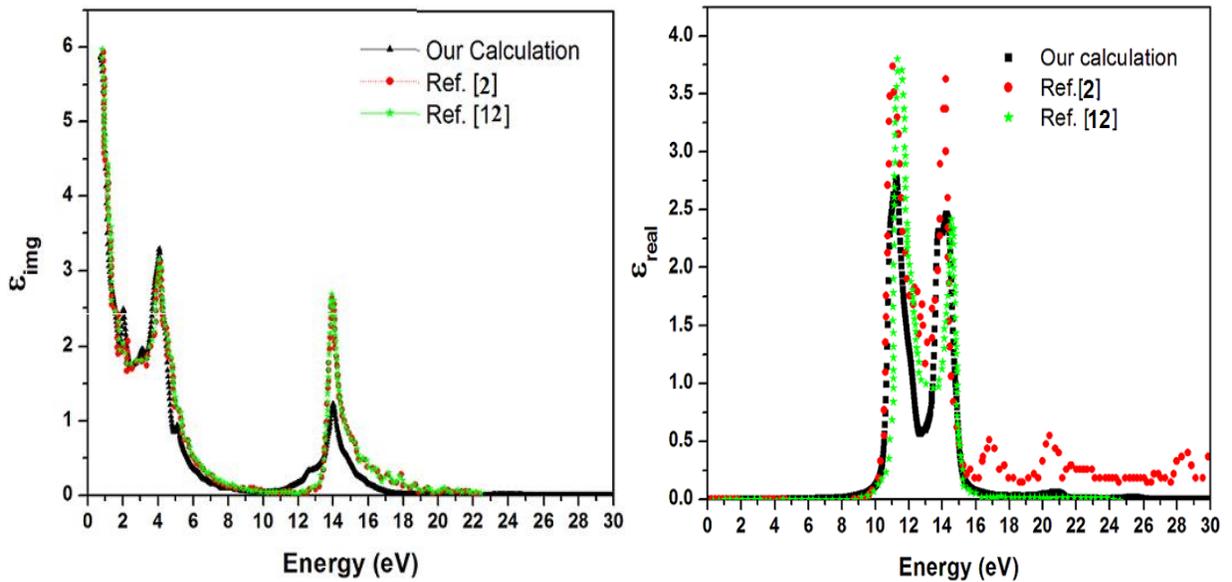

**(a)** **(b)**

**Fig.1. Imaginary part of dielectric function of pure graphene (a) for light polarization parallel to plane of graphene sheet (E⊥c) and (b) light polarization perpendicular to plane of graphene sheet (E∥c) compared with the previous results.**

The absorption spectrum for E⊥$c$ consists of a very significant peak at small frequencies (up to 5 eV) and also another peak structure of broader frequency range which starts from about 10 eV and has a weak intensity peak at 14 eV. The origin of these peak structures is $\pi \rightarrow \pi$* and $\sigma \rightarrow \sigma$* interband transitions, respectively, according to the previous interpretations by Marinopoulos et al. [2]. Our calculations of the absorption spectrum of pure graphene for specific transitions between bands are in good agreement with their results as well as by Sedelnikova et. al [12]. The experimental value of the high intensity peak is 4.6 eV. The lower value of this in our calculations is due to neglect of exchange – correlation within the RPA approach. Imaginary part of dielectric function is shown in Fig. 1(a), has a singularity at zero frequency in E⊥c and thus have optically metallic property but in E∥c shows semiconductor properties.

Now, because of optical selection rules, anisotropy in the optical spectra can be expected i.e. only $\pi \rightarrow \pi^*$ and $\sigma \rightarrow \sigma^*$ transitions are allowed if the light is polarized parallel to the graphene layer, in contrast, only $\pi \rightarrow \sigma^*$ and $\sigma \rightarrow \pi^*$ transitions are allowed if the light is polarized perpendicular to graphene sheet.

The imaginary part of dielectric constant of graphene, for out of plane polarization (E∥c) consists of two prominent peaks at about 11 and 14 eV, as in the earlier calculations by Marniopoulos et. al [13] and Sedelnikova et al. [12]. The spectrum for this polarization differs from that of graphite as graphite has a weak intensity peak in energy range 0-4 eV which is absent here.

The real part of the dielectric function of pure graphene is plotted in fig. 2. Value of static dielectric constant (value of dielectric function at zero energy) in case of E ⊥ c is 7.6 while in case of E∥c is 1.25.

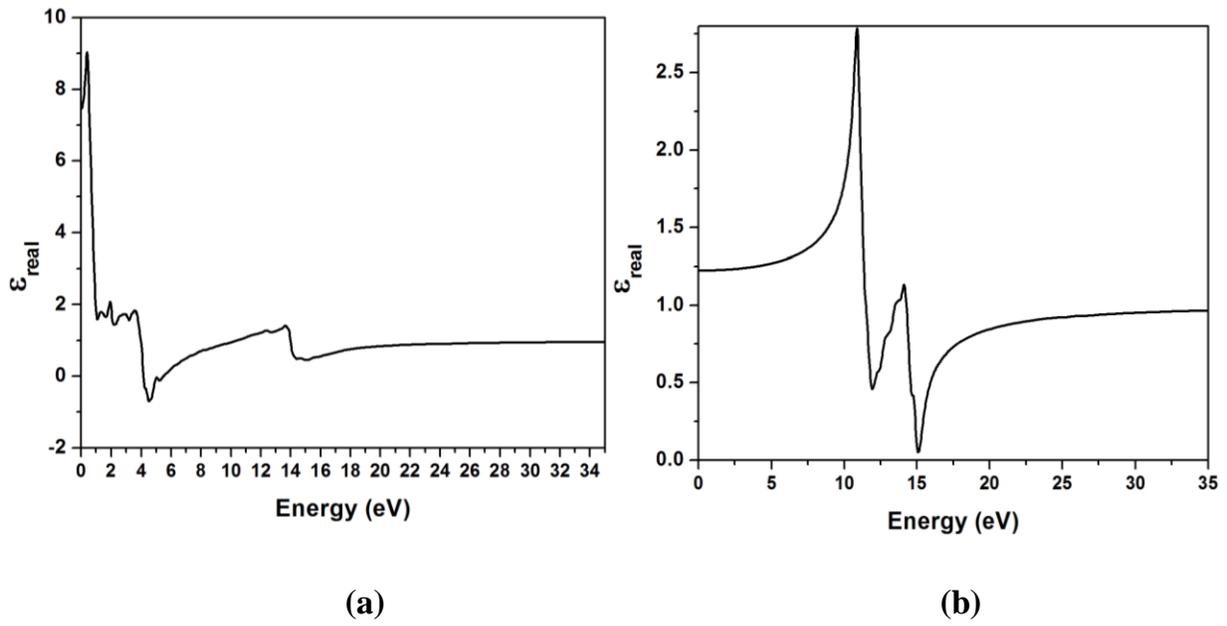

**(a)**          **(b)**

**Fig.2. Real part of dielectric function of pure graphene for *E⊥c (a) and E∥c (b)***

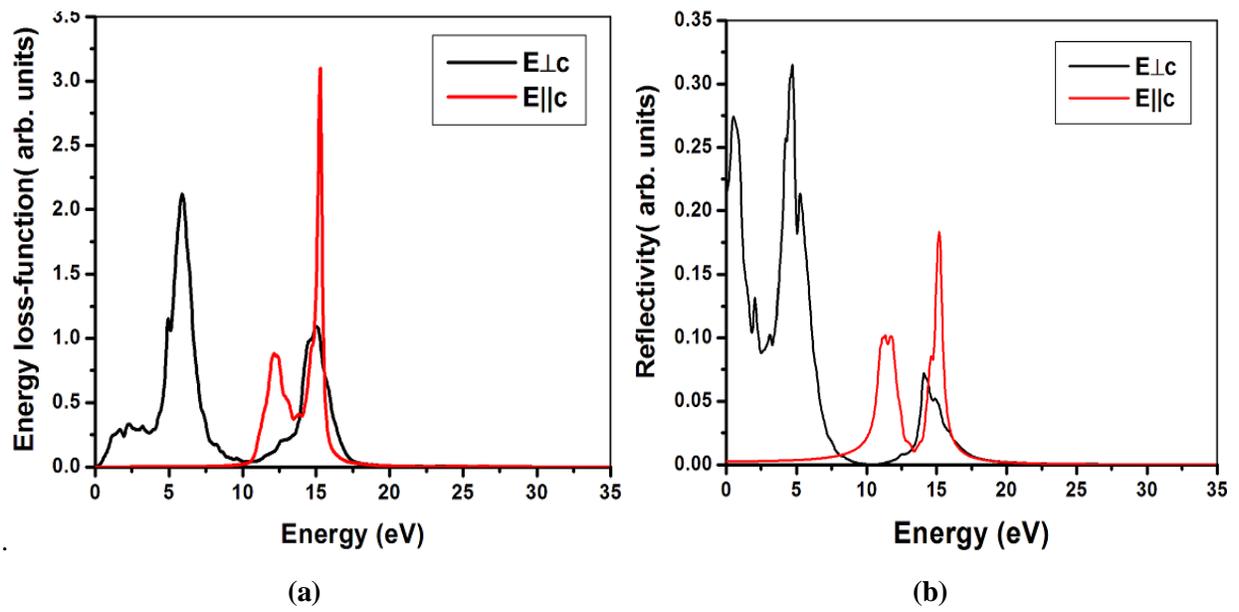

.

**(a)**          **(b)**

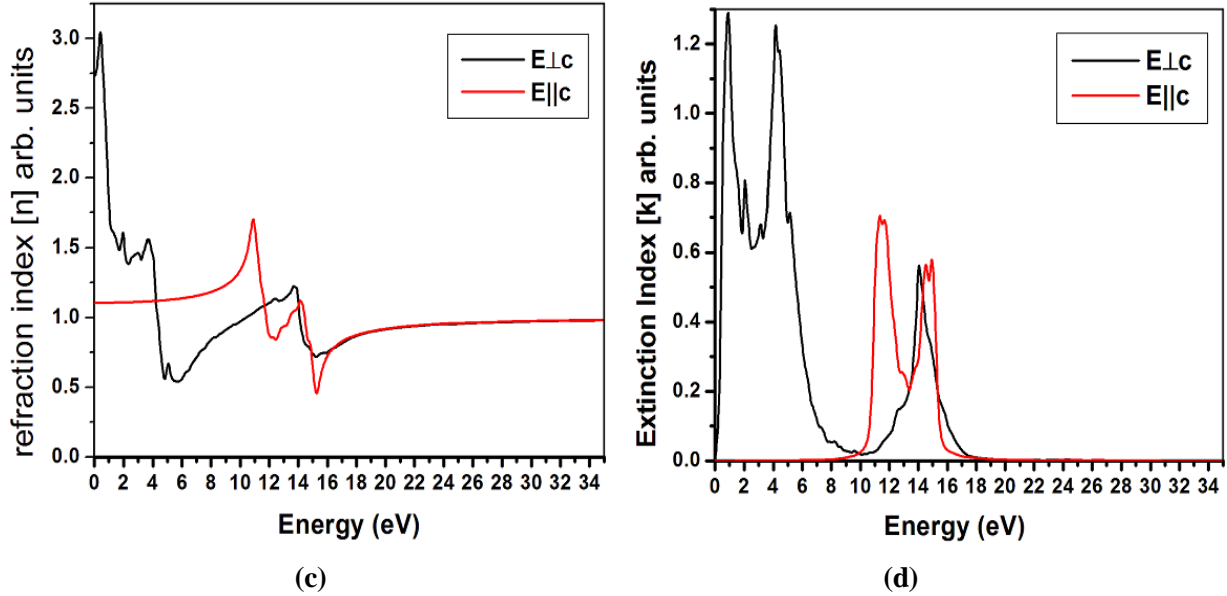

(c)                                                      (d)

**Fig.3. The electron energy loss function (a), reflectivity (b), refraction index (c) and extinction index (d) of pure graphene for E⊥c and E∥c.**

Fig. 3 shows different optical properties of pure graphene. The reflectivity for pure graphene sheet for both directions of electric field is shown in Fig. 3(b). It is found that for parallel polarization of light w.r.t to plane of graphene sheet, reflectivity at lower energy is more and at this energy range, transition is less. Whereas in the light polarization perpendicular to graphene, reflectivity at energy range between 10 and 15eV is more.

The value of static refraction index (value of refraction index at zero energy) in case of E⊥c is 2.75 as shown in Fig. 3 (c), while in case of E∥c is 1.12. Refraction index is minimum at energy of 4.6 eV in (E⊥c) and at 11 eV in E∥c where absorption is maximum.

## 3.2 Effect of doping

After successfully reproducing the results for pure graphene, we employed our method to investigate the effect of hetero-atom doping on the optical properties of graphene. The doped graphene sheets were theoretically generated and C atoms were substituted with both individual B and N and then B/N. The concentration of individual B or ( N) dopants were varied from 3.125 % (one atom of the dopant in 32 host atoms) to 6.25 % (six dopant atoms in 50 host atoms) and from 37.5 % (one B/N pair in 32 host atoms) to 18.75 % for BN co-doping. Positions of the dopants have also been varied for the same concentration of substitution doping to analyze the effect of isomerization.

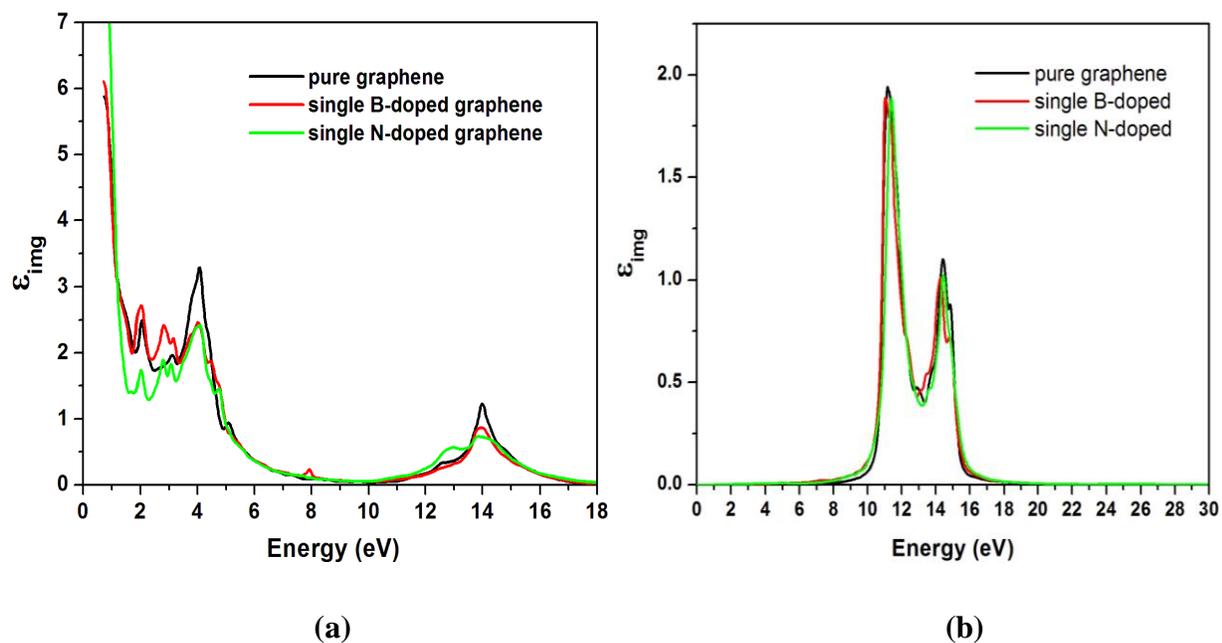

**(a)** **(b)**

**Fig.4. Comparison of imaginary part of dielectric function of pure graphene with that of single boron and nitrogen atom doped graphene sheet for E⊥c (a) and E||c (b).**

In Fig. 4 the imaginary part of dielectric function of individual single B and N doped graphene are plotted in comparison to the pure graphene. From the results for perpendicular polarization of light (Fig.4(a)) , we can see that the 4 eV peak in single B doped and single N doped graphene sheet is of lesser intensity and the well- defined peak changes into a broader plateau. Since on individual B and N doping of graphene, only Fermi level shifts and band gap is not introduced at Fermi level so energy peak at 4.0 eV is not shifted.

Also it can be inferred from results for parallel polarization of light as shown in fig. 4(b) that doping with single B and N atom has almost no effect on the dielectric constant for out of light polarization. Because as pointed out by Marinopoulos et al. [2] for this light polarization the band structure does not play the exclusive role in defining the absorption spectrum. Only local field effects affect the peak positions which are not taken into account in RPA approximation.

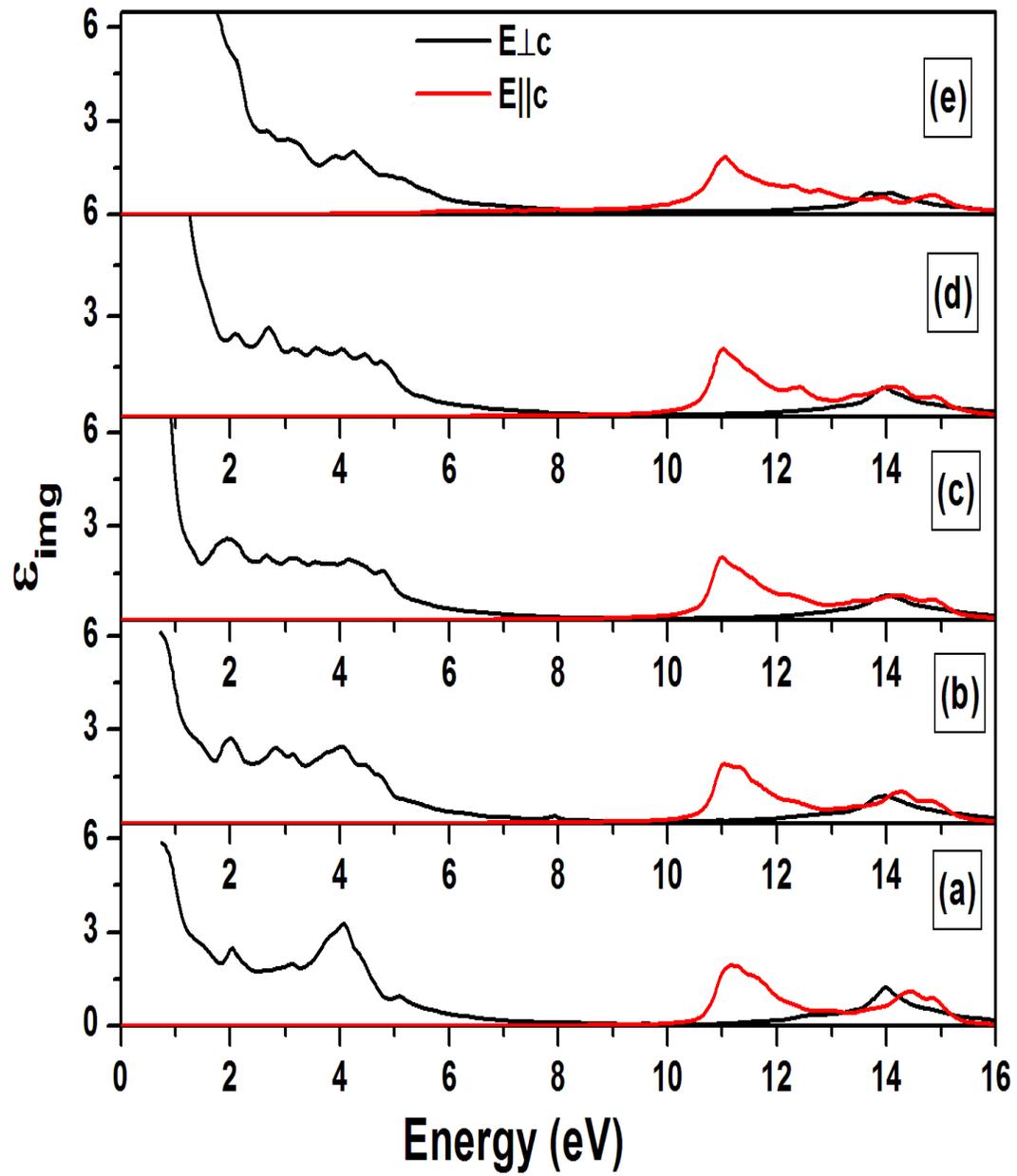

(a)

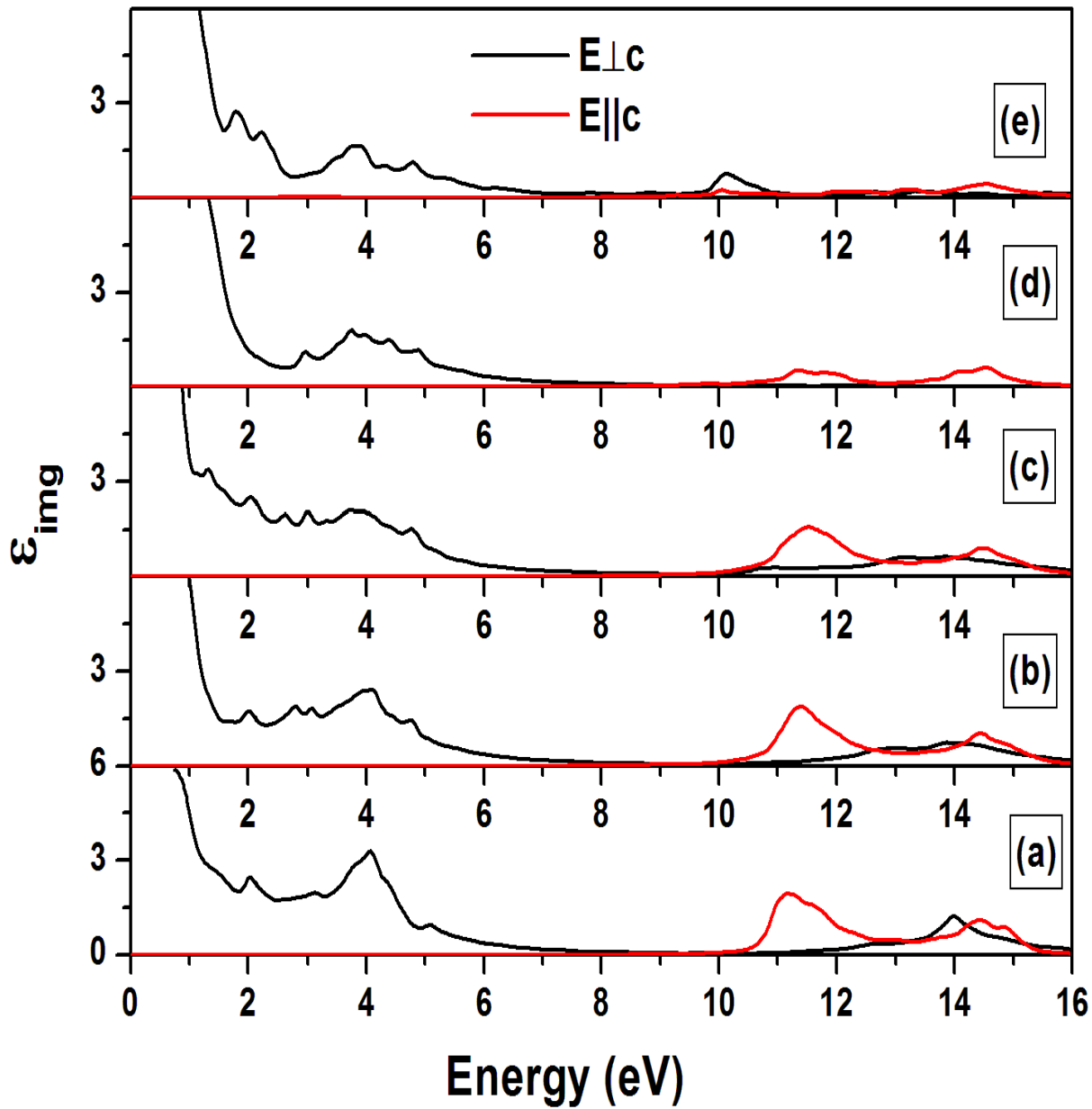

**(b)**

**Fig 5(b). The imaginary part of dielectric function for pure graphene (a) as compared with graphene sheet doped with different concentrations of boron (a) and nitrogen (b) in increasing order, 3.125% (b), 6.25% (c), 9.375% (d) and 18.75% (e) respectively for E⊥c and E∥c.**

The results [Fig. 5(a).] show that as we go on increasing the B-doping concentration, the intensity of the main absorption peak goes on decreasing regularly. At very higher concentration of doping at about (18.75%) the high intensity peak at 4eV completely disappears indicating that there is no absorption. The peak position and intensity for out of plane polarization of light does not change as in the case of single B-doping. So the 4eV peak is of main consideration for tailoring the optical properties of graphene in visible range of light.

The main trends in effects on the imaginary part of dielectric function of the N-doping [fig. 5 (b)] are same as that of B-doping. Only the decrease in intensity of the peak is more rapid and prominent.

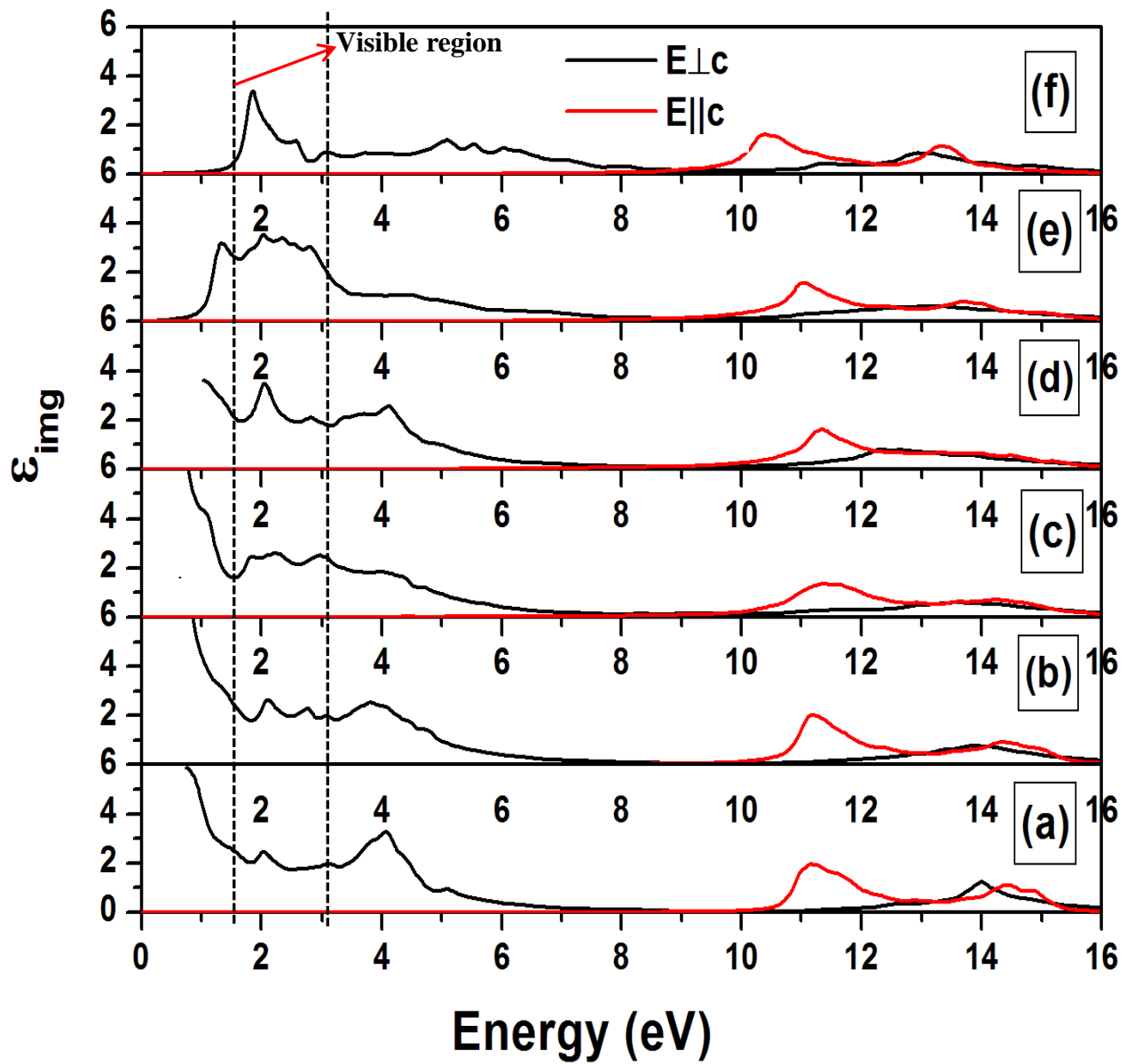

**Fig 6. The imaginary dielectric function for pure graphene (a) as compared with graphene sheet co-doped with different concentrations of BN in increasing order concentrations of boron in increasing order, 6.25% (b), 12.5% (c), 18.75% (d) and 37.5% (e) and 75% respectively for E⊥c and E∥c.**

The comparison of imaginary part of dielectric function of pure graphne as compared to the B/N co-doped graphene at diiferent concentrations is given in Fig. 6. On BN codoping the intensity of the 4 eV peak goes on decreasing and is red shifted towards 3.8 eV at 6.25% doping concentartion. With further increase in the doping , the  intensity of this peak goes on decreasing and a peak in the visible range starts to appear. At very high concentration of doping, a new peak at 2 eV ( in the visible region) is identified. We tabulate the salient features highlighting differences in shifts due to doping in Table 1 also.

**Table 1. Main absorption peak position and shift in peak position ($\delta E$ ) due to doping**

| Doping conc. (%) | B doping | | N doping | | Doping conc. (%) | B/N co-doping | |
|---|---|---|---|---|---|---|---|
| | Peak Position | ΔE | Peak Position | ΔE | | Peak Position | ΔE |
| 0 (pure graphene) | 4.0 | - | 4.0 | - | 0  ( pure graphne) | 4.0 | - |
| 3.125 | 4.0 | 0 | 4.0 | 0.0 | 6.25 | 3.8 | 0.2 |
| 6.25 | 3.9 | | 3.8 | 0.2 | 12.5 | 3.3 | 0.7 |
| 9.375 | 3.8 | | 3.8 | 0.2 | 18.75 | 2.8 | 1.2 |
| 18.75 | Peak disappears | | 3.8 | - | 37.5 | 2.5 | 1.5 |
| | | | | | 75 | 1.9 | 2.1 |

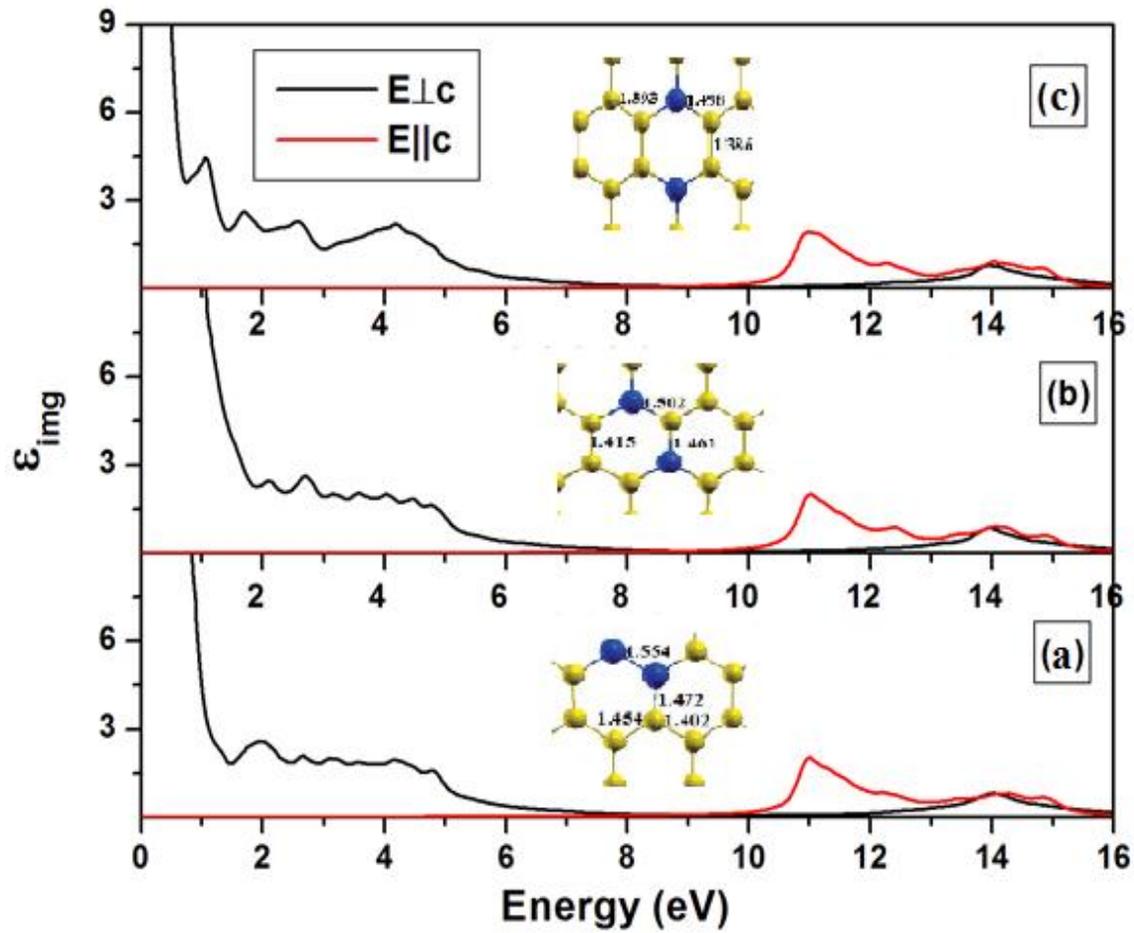

**Fig.7 (a). Comparison of imaginary part of dielectric function of three different isomers of graphene sheet doped with 2 boron atoms with configurations as shown in inset for in–plane light polarization (E⊥c).**

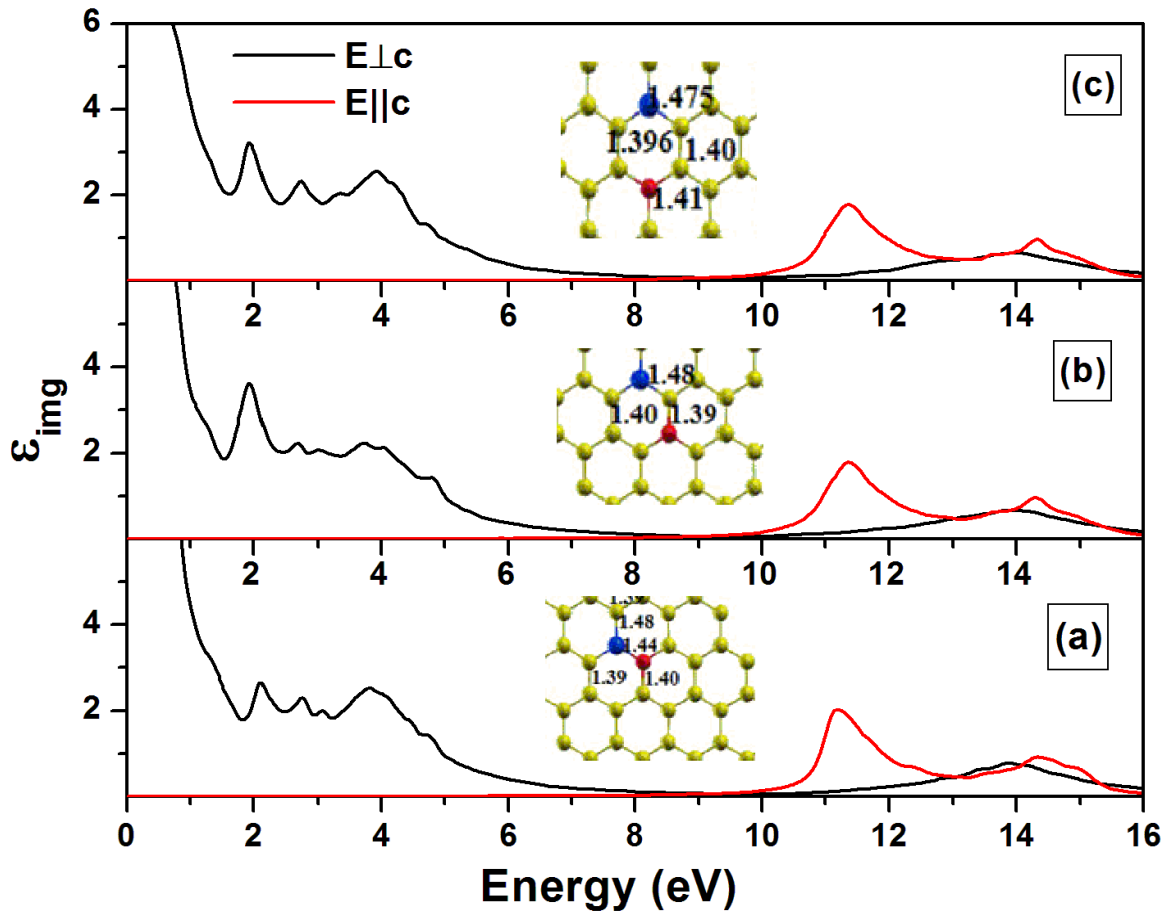

**Fig.7 (b). Comparison of imaginary part of dielectric function of pure graphene with that of B/N co-doped graphene sheet for different doping sites, ortho (a), meta (b) and para isomer (c).**

If we compare the imaginary part of dielectric function of different isomers, the absorption peaks completely vanish in the isomer showing maximum band gap in case of B doping (Fig. 7(a)), but in case of BN co-doping a new peak in the visible region appears more predominant in para isomer (Fig. 7(b))

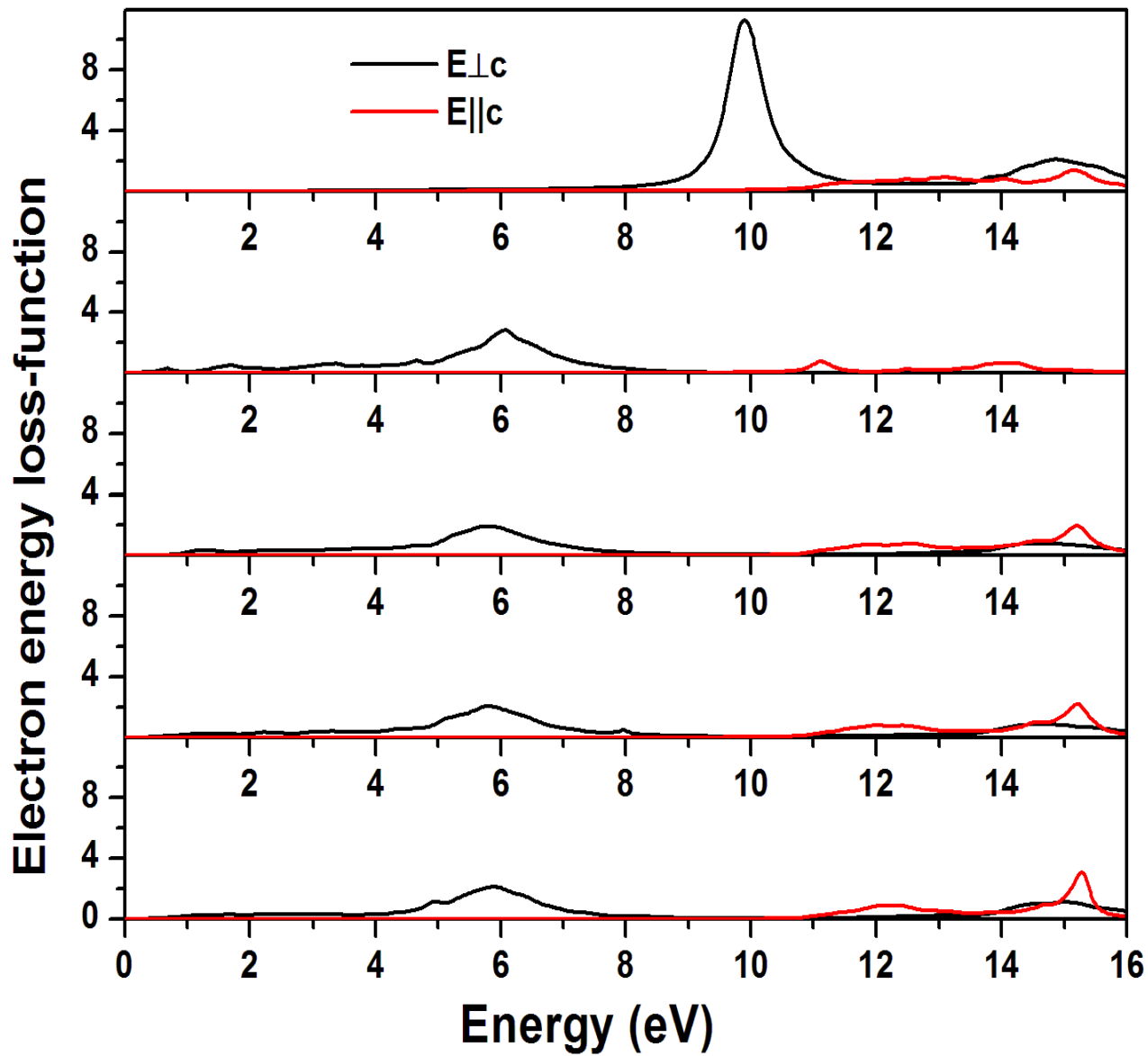

**Fig 8 (a). The energy-loss function for pure graphene (a) as compared with graphene sheet doped with different concentrations of boron in increasing order (upwards), 3.125% (b), 6.25% (c), 9.375% (d) and 18.75% (e) respectively.**

It can be seen from the plots of electron energy loss function (fig. 8.) that with increasing concentration of boron doping the intensity of the 15 eV peak goes on decreasing and intensity of the 5.8 eV peak goes on increasing (fig. 8(a)). At very high concentration of doping (18.75%) the 5.8 eV peak blue shifts to 10 eV.

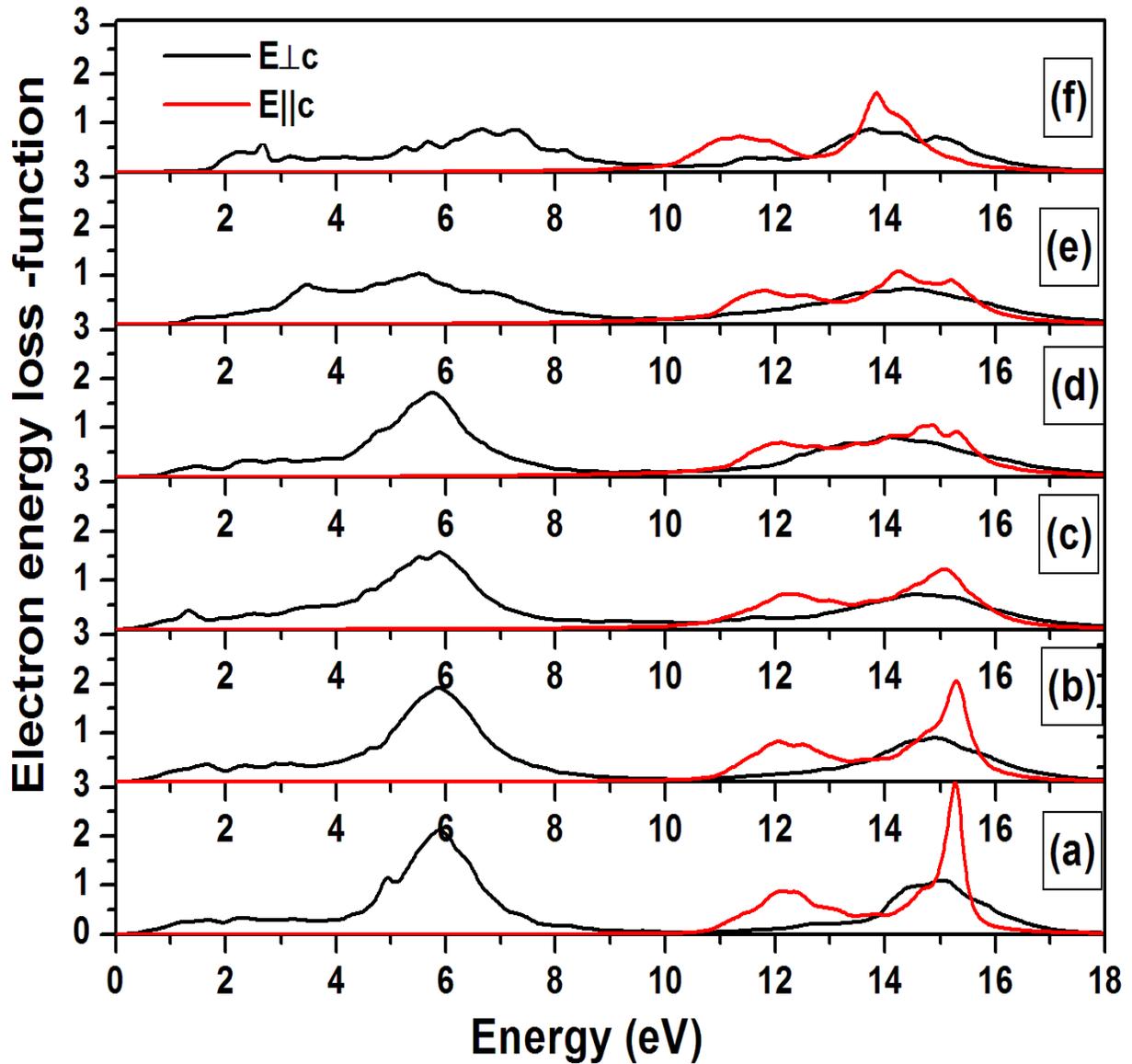

**Fig 8 (b). The electron –energy loss spectra for pure graphene (a) as compared with graphene sheet co-doped with different concentrations of BN in increasing order concentrations of boron in increasing order, 6.25% (b), 12.5% (c), 18.75% (d) and 37.5% (e) and 75% respectively for E⊥c and E∥c.**

Fig 8 (b). represents the comparison of the EELs function of pure graphene with that of  B/N co-doped graphene at different concentrations.

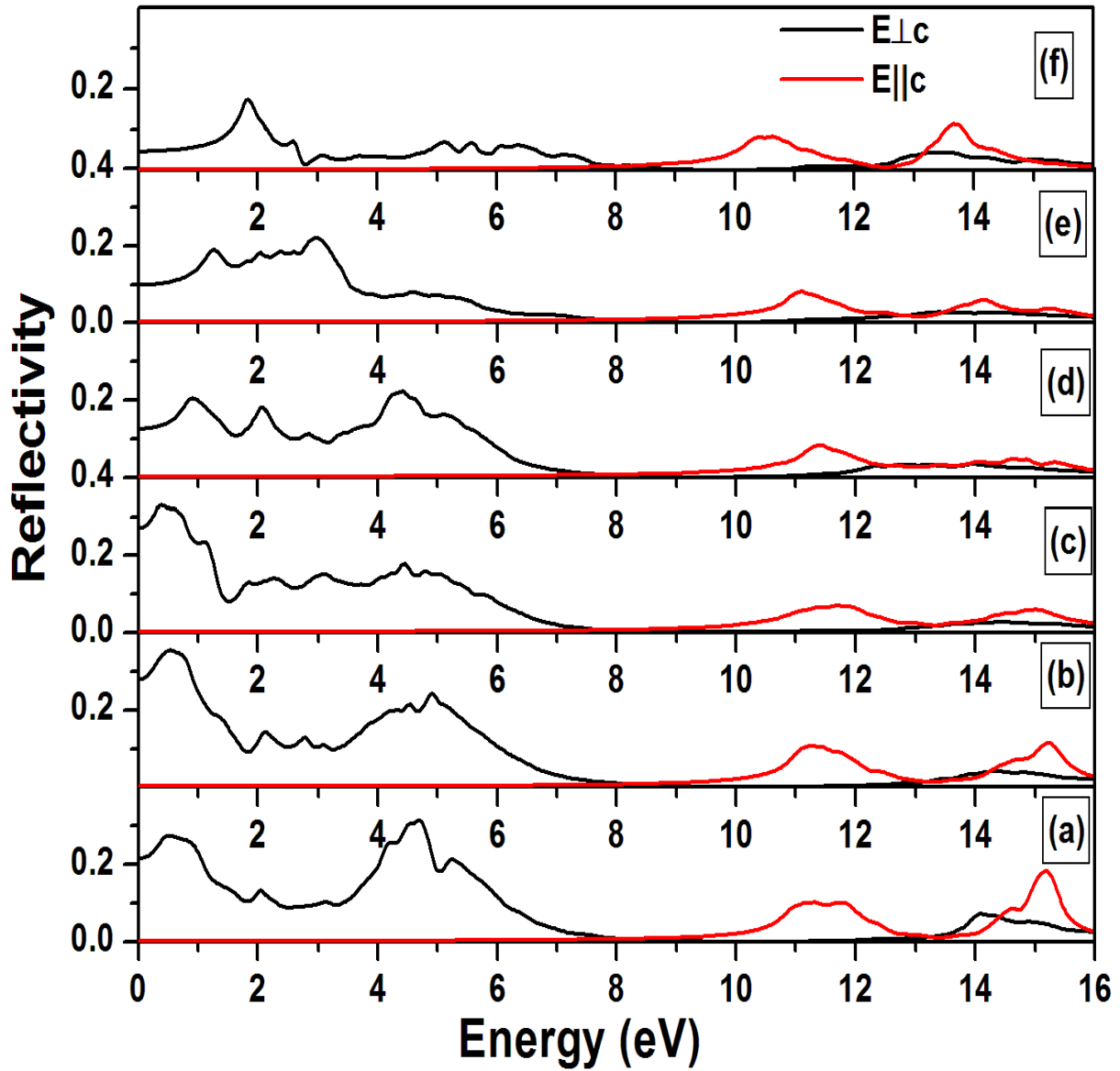

**Fig 9. The reflectivity for pure graphene (a) as compared with graphene sheet co-doped with different concentrations of BN in increasing order concentrations of boron in increasing order, 6.25% (b), 12.5% (c), 18.75% (d) and 37.5% (e) and 75% respectively for E⊥c and E∥c.**

Fig.9 gives the comparison of reflectivity of pure graphene with that of B/N co-doped graphene at different concentrations.

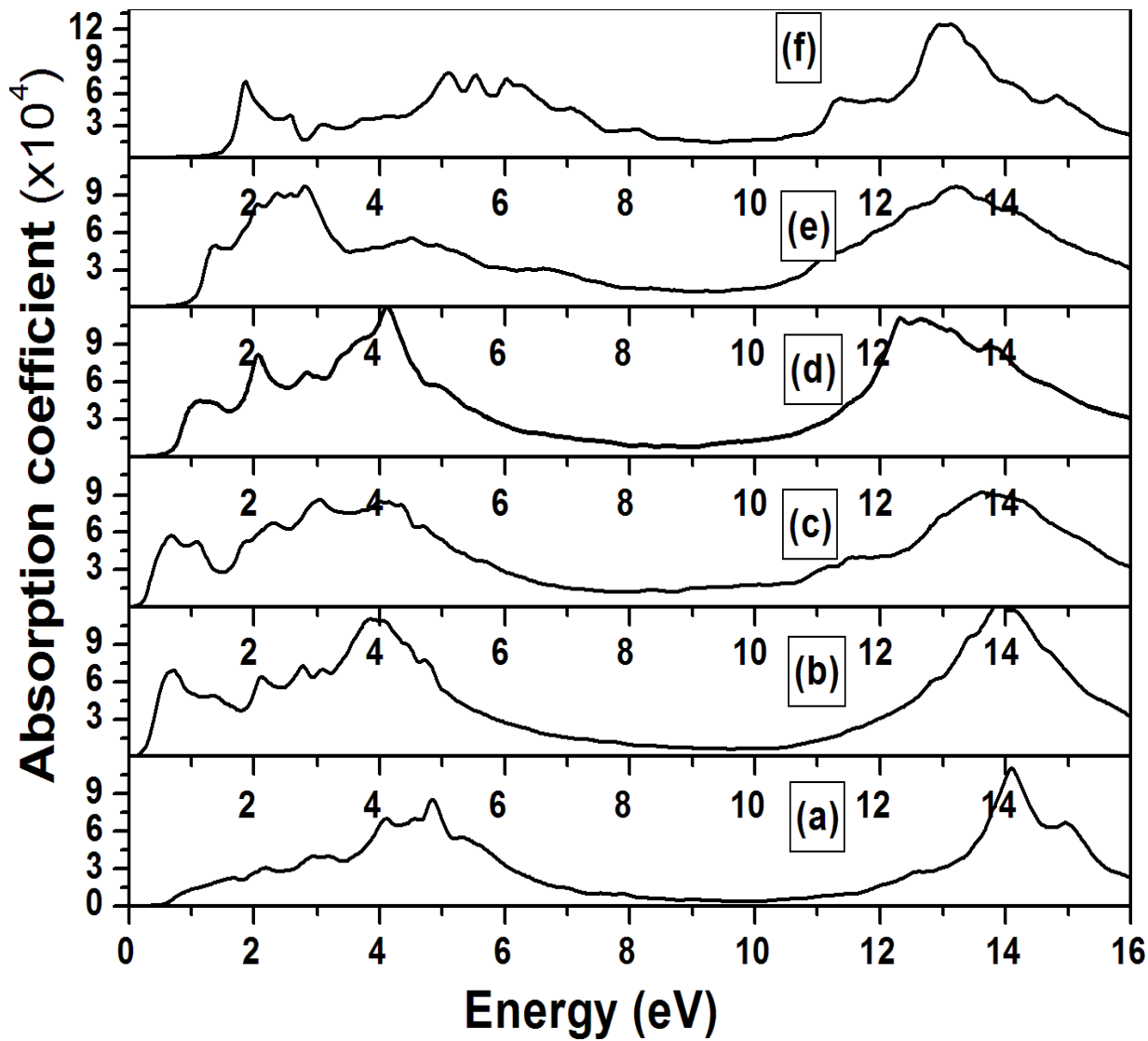

**Fig 10. The absorption coefficient for pure graphene (a) as compared with graphene sheet co-doped with different concentrations of BN in increasing order concentrations of boron in increasing order, 6.25% (b), 12.5% (c), 18.75% (d) and 37.5% (e) and 75% respectively for E⊥c.**

The plots for the absorption coefficient for pure graphene in comparison to the B/N co-doped graphne are shown in Fig. 10. Like the graphs for imaginary part of dielectric constant of garphene, the absorption coefficient alos shows peaks at about 4eV and 14 eV. At about 75% doping concentration, the absorption peak appears in visible region indicating the absorption in visible region.

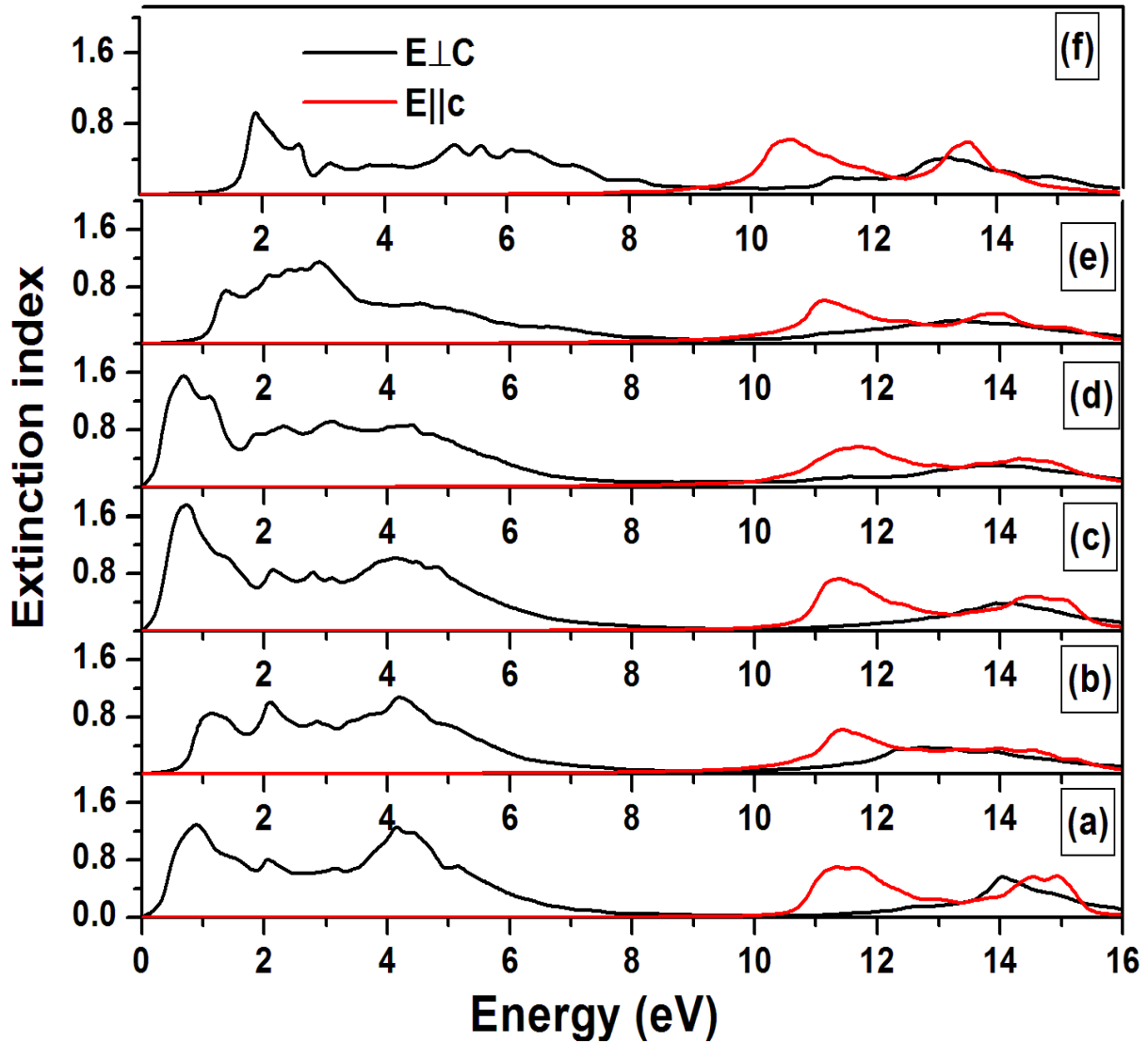

**Fig 11. The extinction index for pure graphene (a) as compared with graphene sheet co-doped with different concentrations of BN in increasing order concentrations of boron in increasing order, 6.25% (b), 12.5% (c), 18.75% (d) and 37.5% (e) and 75% respectively for E⊥c and E||c.**

Fig. 11 shows the extinction coefficient of graphene sheet doped with different concentrations of B/N as compared to the pristine graphene. As shown in the plot, the extinction index shows a peak in visible region at very high concentration of B/N co-doping.

## 4. Summary and Conclusions

DFT within RPA approximation was employed to calculate the optical properties of pure and doped graphene. The dielectric function, absorption spectrum, reflectivity and energy loss-function of single layer graphene sheet have been calculated for light polarization parallel and perpendicular to the plane of graphene sheet and compared with doped graphene. The calculated dielectric functions, energy-loss spectra and other properties of pure graphene were found to be fairly in good agreement with the available data.

The doping by B or N or a combination of B and N by altering the concentration of B/N as well as position of doping has been extensively studied. The results show that with individual B (or N) doping, the intensity of the absorption peaks goes on decreasing with increasing concentration and the main absorption peak (at about 4eV) almost vanishes at about 18.75% doping level. But for this type of doping there is almost no or little shift in the main absorption peak. So Individual B (or N) doping is not very useful if we want to tailor the optical properties of graphene in visible region.

On B/N co-doping, the main absorption peak continuously goes on red-shifting with increasing concentration from the original position of 4eV. The peak appears completely in visible region (about 6525 Å) at a very high doping level of 75%. This pattern is also observed in case of reflectivity and absorption coefficient.

In conclusion the individual B and N doping does not significantly affect the imaginary dielectric function and hence the absorption spectra. However, significant red shift in absorption towards visible range of the light is found to occur in case of B/N co-doping at high doping concentration. We can conclude that the B/N co-doping of graphene can alter the optical properties of graphene to make it reflect in the visible region. It thus seems that appropriate concentration and positional doping of a combination of B/N helps in significant modification of the absorption spectra of graphene. Our findings suggest further experimental investigations in this regard which could lead to application of graphene in field of photonics where absorption in visible region is required.

## Acknowledgements

We express our gratitude to VASP team for providing the code, the HPC facilities at IUAC (New Delhi) and the departmental computing facilities at Department of Physics, PU, Chandigarh. PR gratefully acknowledges financial support from UGC. VKJ acknowledges support from CSIR as Emeritus Scientist.